\documentclass[twocolumn, prl, amssymb, superscriptaddress, aps, preprintnumbers,
amsmath,showkeys,floatfix]{revtex4}

\usepackage{epstopdf}
\usepackage{graphics}
\usepackage{graphicx}
\usepackage{dcolumn}
\usepackage{bm}
\usepackage{longtable}
\usepackage{epsfig}
\usepackage{times}
\usepackage{url}
\usepackage{color}

\begin{document}
%%%%%%%%%%%%%%%%%%%%%%%%%%%%%%%%%%%%%%%%%%%%%%%%%%%%%%%%%%%%%%%%%%%%%%%%%%%%%%%%
%%%%%%%%%%%%%%%%%%%%%%%%%%%%%%%%%%%%%%%%%%%%%%%%%%%%%%%%%%%%%%%%%%%%%%%%%%%%%%%%
\title{Spectral control over $\gamma$-ray echo using a nuclear frequency comb system}
 
\author{Chia-Jung   \surname{Yeh}}
\affiliation{Department of Physics, National Central University, Taoyuan City 32001, Taiwan}
\author{Po-Han   \surname{Lin}}
\affiliation{Department of Physics, National Central University, Taoyuan City 32001, Taiwan}
\author{Xiwen    \surname{Zhang}}
\affiliation{Department of Physics and Astronomy, University of Toronto, Toronto, Ontario M5S 1A7, Canada}
\author{Olga  \surname{Kocharovskaya}}
\affiliation{Department of Physics and Astronomy, Texas A\&M University, College Station, Texas 77843, USA}
\author{Wen-Te \surname{Liao}}
\email{wente.liao@g.ncu.edu.tw}
\affiliation{Department of Physics, National Central University, Taoyuan City 32001, Taiwan}

\date{\today}
%%%%%%%%%%%%%%%%%%%%%%%%%%%%%%%%%%%%%%%%%%%%%%%%%%%%%%%%%%%%%%%%%%%%%%%%%%%%%%%%
%%%%%%%%%%%%%%%%%%%%%%%%%%%%%%%%%%%%%%%%%%%%%%%%%%%%%%%%%%%%%%%%%%%%%%%%%%%%%%%%
\begin{abstract}
Two kinds of spectral control over $\gamma$-ray echo using a nuclear frequency comb system are theoretically investigated. A nuclear frequency comb system is composed of multiple  nuclear targets under  magnetization (hyperfine splitting),  mechanical motion (Doppler shift) or both, namely, moving and magnetized targets. In frequency domain the unperturbed single absorption line of $\gamma$-ray  therefore splits into multiple lines with equal spacing and becomes a nuclear frequency comb structure. We introduce spectral shaping and dynamical splitting to the frequency comb structure respectively to  optimize the use of a medium and to break the theoretical maximum of echo efficiency, i.e., 54\%. 
Spectral shaping scheme leads to the reduction of required sample resonant thickness for achieving high echo efficiency of especially a broadband input.
Dynamical splitting method significantly advances the echo efficiency up to 67\% revealed by two equivalent  nuclear frequency comb systems.
We also show that using only few targets is enough to obtain good echo performance, which significantly eases  the complexity of implementation. Our results extend  quantum optics to 10keV regime and lay the foundation of the development of $\gamma$-ray memory. 
%Moreover, the present echo scheme suggests an application of having tunable time delay for a $\gamma$-ray pulse via the modulation of nuclear frequency comb.
\end{abstract}
%%%%%%%%%%%%%%%%%%%%%%%%%%%%%%%%%%%%%%%%%%%%%%%%%%%%%%%%%%%%%%%%%%%%%%%%%%%%%%%%
%\pacs{
%42.50.-p, %Quantum optics
%}

%\keywords{quantum optics}
%%%%%%%%%%%%%%%%%%%%%%%%%%%%%%%%%%%%%%%%%%%%%%%%%%%%%%%%%%%%%%%%%%%%%%%%%%%%%%%%
\maketitle
%%%%%%%%%%%%%%%%%%%%%%%%%%%%%%%%%%%%%%%%%%%%%%%%%%%%%%%%%%%%%%%%%%%%%%%%%%%%%%%%
%-----------Text body-----------------------------------------------------------
%
%
The $\gamma$-ray-nuclear interfaces are recently demonstrated to be interesting for fundamental physics in the development of $\gamma$-ray/X-ray quantum optics \cite{Adams2013}, which explores a completely new energy realm of light-matter interactions \cite{Burck1987, Kocharovskaya1999, Kuznetsova2003, Buervenich2006, Vagizov2009, Roehlsberger2010, Liao2011, Roehlsberger2012, Liao2013, Heeg2013, Heeg2015a, Radeonychev2015, Liao2016, Haber2017, Khairulin2018}. 
Pursuing coherent control over $\gamma$-rays  is also expected to be versatile for applications \cite{Vagizov2014, Liao2014a, Liao2015, Heeg2015b, Heeg2017, Liao2017, Wang2018}.
E.g., spectral control over $\gamma$-rays may lead to new light source for M\"ossbauer spectroscopy \cite{Heeg2017}.
Given the sub-angstrom wavelength of $\gamma$-photon, it avoids the bottleneck of diffraction limit  for  photonic information processing \cite{Liao2012a, Vagizov2014}, e.g., $\gamma$-photon storage \cite{Shvydko1996, Liao2012a, Kong2016, Zhang2018}. 
$\gamma$-ray echo \cite{Helisto1991, Helisto2001} is shown to be a valuable and practical method for shaping $\gamma$-ray wavepacket \cite{Vagizov2014} and forming nuclear quantum memory \cite{Zhang2018}.
In this letter, we put forward spectral shaping and dynamical splitting to manipulate  the  recently proposed nuclear frequency comb (NFC) \cite{Zhang2018} and its resulting $\gamma$-ray echo.
The former is used to optimize the echo efficiency, and the latter is intended for beating the  theoretically maximum echo efficiency of 54\% \cite{Sangouard2007, Zhang2018}.
In view of the success of the optical echo memory \cite{Sangouard2007, Moiseev2008, Hetet2008, Tittel2010, Zhang2014, Liao2014b, Su2017} based on atomic frequency comb (AFC) \cite{Afzelius2009, Zhou2012}, our scheme gives an importnat step for the development of future $\gamma$-ray echo memory \cite{Zhang2018}. Moreover, the present results suggest applications such as a $\gamma$-ray coherent delay line providing an input $\gamma$-ray pulse with a tunable time delay, and the generation of time-bin qubit for single $\gamma$ quanta.

%%%%%%%%%%%%%%%%%%%%%%%%%%%%%%%%%%%%%%%%%%%%%%%%%%%%%%%%%%%%%%%%%%%%%%%%%
\begin{figure}[b]
\vspace{-0.4cm}
  \includegraphics[width=0.45\textwidth]{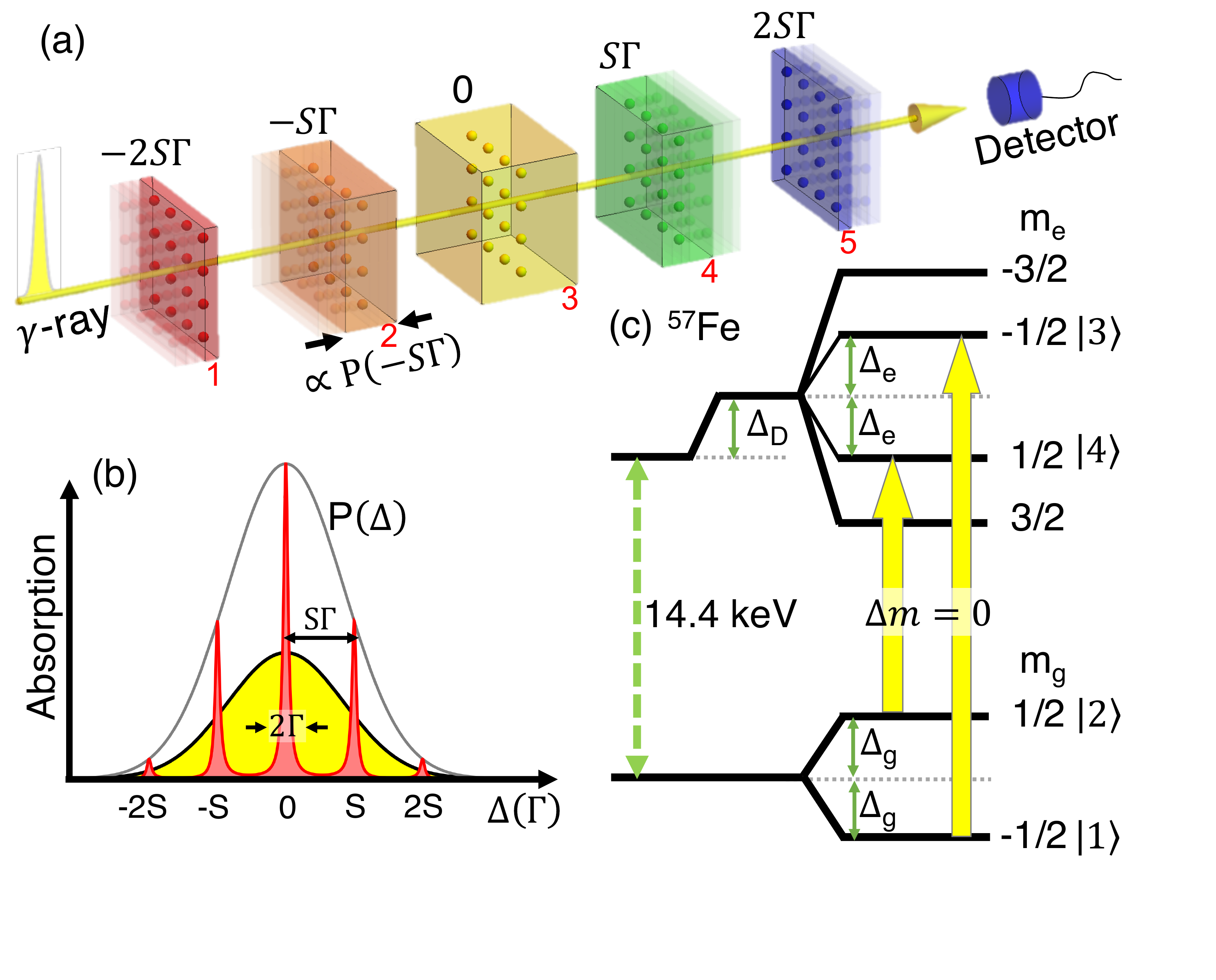}
  \caption{\label{fig1}
(Color online) 
(a) nuclear frequency comb system made of multiple targets with a resonant thickness distribution $P\left( \Delta\right)$. Cuboid of each color depicts nuclei in which experience an energy shift specified above cuboid, e.g., red for a shift of $-2S\Gamma$ from unperturbed transition energy.
(b) spectral shaping of a nuclear frequency comb by introducing $P\left( \Delta\right)$.
Nuclear frequency comb with equal spacing $S\Gamma$ formed by nuclear absorption lines of different targets, and each linewidth is $2\Gamma$. (c) $^{57}$Fe nuclear level scheme. A linearly polarized $\gamma$-ray (yellow-upward-wide arrows) drives $\Delta m=0$ transitions.  $\Delta_g$ ($\Delta_e$) is the ground (excited) state hyperfine splitting, and $\Delta_D$ is the Doppler shift. The situation when both $\Delta_D\neq 0$ and $\Delta_g+\Delta_e\neq 0$ is termed as hybrid shift. 
}
\end{figure}
%%%%%%%%%%%%%%%%%%%%%%%%%%%%%%%%%%%%%%%%%%%%%%%%%%%%%%%%%%%%%%%%%%%%%%%%%
%
We begin with the optimization of a NFC using spectral shaping.
Figure \ref{fig1}(a) illustrates our modified NFC system by  introducing a resonant thickness distribution function $P\left( \Delta \right)$ to a sequence of targets.
%, i.e., each target has a corresponding product of thickness and particle concentration proportional to $P\left( \Delta \right)$. 
%
The spatial order of the comb teeth does not matter.
The energy shift $\Delta$  from the unperturbed transition energy can be generated by, e.g., Doppler shift, and the afterimage of each cuboid depicts its motion with a constant velocity \cite{Zhang2018}. 
%The nuclear transition energy of each sample is different from others due to magnetization (hyperfine splitting) \cite{Shvydko1996, Liao2012a} or mechanical motion (Doppler shift) \cite{Ruby1960, Helisto1991, Vagizov2014, Zhang2018}. 
%
The coloured cuboids represent targets where embedded nuclei experience different $\Delta$.
With a proper arrangement, multiple absorption peaks can form a NFC with a tunable comb spacing $S\Gamma$. Here $\Gamma$ is the spontaneous decay rate of some chosen excited states, and $S>0$.
Red, orange, yellow, green and blue are respectively for  $\Delta = -2S\Gamma$,  $-S\Gamma$,  zero,  $S\Gamma$ and  $2S\Gamma$.
The yellow short pulse depicts the incident $\gamma$-ray.
Since the absorption strength is proportional to resonant thickness, $P\left( \Delta \right)$ modifies the overall NFC in the frequency domain  as depicted in Fig.~\ref{fig1}(b).
The black-solid-yellow-filled line demonstrates the spectrum of the incident $\gamma$-ray pulse.
In view that the input spectral intensity 
on resonance with unperturbed transition is stronger than those at side bands, $P\left( \Delta \right)$ can properly match the overall NFC (multiple red-solid-filled peaks)  with the input spectrum such that the use of whole medium is optimal.  
This will ease the demand of very high resonant thickness for achieving high echo efficiency in a flat comb \cite{Zhang2018}.

Figure~\ref{fig1}(c) shows the level scheme of $^{57}$Fe nucleus with hyperfine splitting $\Delta_g+\Delta_e$ and Doppler shift $\Delta_D$.
Without Doppler shift, i.e., $\Delta_D=0$, multiple magnetized and stationary targets can also form a NFC when $\Delta_g+\Delta_e =S\Gamma$, and each target contributes a pair of comb tooth for a linearly polarized $\gamma$-ray.
For the sake of simplicity we term the case when both $\Delta_D\neq 0$ and $\Delta_g+\Delta_e\neq 0$ as hybrid shift. 
The Maxwell-Bloch equation for  the $n$th target in perturbation regime \cite{Scully2006, Liao2012a, Liao2014b, Kong2014}, namely, $\vert\Omega_n\vert\ll\Gamma$, describes the coupling between nuclei and $\gamma$-ray:
\begin{eqnarray}
\partial_t \rho_{31}^n &=& -\left[ \frac{\Gamma}{2}+i \Delta_{31}^n\left( t\right) \right] \rho_{31}^n+i\frac{a}{4}\Omega_n, \label{eq1} \\
\partial_t \rho_{42}^n &=& -\left[ \frac{\Gamma}{2}+i \Delta_{42}^n\left( t\right) \right] \rho_{42}^n+i\frac{a}{4}\Omega_n, \label{eq2}\\
\frac{1}{c}\partial_t \Omega_n &+& \partial_z \Omega_n =i\eta_n a\left(\rho_{31}^n+\rho_{42}^n\right) \label{eq3}.
\end{eqnarray}
Together with initial conditions
$\rho_{31}^n\left( 0, z\right)=\rho_{42}^n\left( 0, z\right)=0$, $\Omega_n \left( 0, z\right)=0$ and the input boundary condition 
$\Omega_1 \left( t, 0\right)=\Omega_0 Exp\left[ -\left( t-\tau_i \right)^2/ \tau_p^2  \right]$. Other boundary conditions are $\Omega_{n>1} \left( t, 0\right)=\Omega_{n-1} \left( t, L_{n-1}\right)$, i.e., we treat the output from $\left( n-1\right)$th target as the input of $n$th target.
Here $\rho_{31}^n$ and $\rho_{42}^n$ are the slowly varying amplitudes of the coherence term in the density matrix of the level scheme illustrated in Fig.~\ref{fig1}(c). 
$\Omega_n$ is the slowly varying amplitudes of the  Rabi frequency of the linearly polarized $\gamma$-ray propagating through $n$th target, and $c$ represents the speed of light.
Above mentioned $\Delta$ is generalized for $^{57}$Fe nuclear hyperfine structure.
Energy shift of transition $\vert 1 \rangle\rightarrow\vert 3 \rangle$ and $\vert 2 \rangle\rightarrow\vert 4 \rangle$ are respectively
$\Delta_{31}^n=\Delta_g^n+\Delta_e^n+\Delta_D^n$ and $\Delta_{42}^n=-\left( \Delta_g^n+\Delta_e^n\right) +\Delta_D^n$.
The coupling constant  $\eta_n =  6\Gamma\xi_n / L_n  $, where $\Gamma = 1/141.1$ GHz is the  spontaneous decay rate of excited states $\vert 3 \rangle$ and $\vert 4 \rangle$, $\xi_n$ the resonant thickness and $L_n$ the thickness of $n$th target. 
$a = \sqrt{2/3}$ is the corresponding Clebsch-Gordan coefficient  for two $\Delta m = m_e - m_g = 0$ transitions in Fig.~\ref{fig1}(c).
For flat NFC, a Fourier analysis \cite{Zhang2016, Zhang2018} of Eq.~(\ref{eq1}-\ref{eq3}) shows that the echo efficiency 
$\mathbb{E}=
\left[ \int_{t_1}^{t_2}\vert \Omega_M\left( t, L_M\right) \vert^2 dt\right]  /
\left[ \int_{-\infty}^{\infty}\vert \Omega_1\left( t, 0\right) \vert^2 dt\right]  = 
16\pi^2 \overline{\xi}^2 \mathrm{Exp}\left[ -2\pi\left( 2\overline{\xi}+1\right)/S  \right]/
S^2
$
where $\overline{\xi} = \sum_{n=1}^M \xi_n/M$ is the average resonant thickness over total $M$ targets. Temporal domain $\left[t_1, t_2 \right]$ contains only echo signal which peaks at $\tau_e=\tau_i+2\pi/\left( S\Gamma\right) $. The theoretical maximum $\mathbb{E}$ is 54\%  when $\overline{\xi}$ and $S$ approach infinity as also predicted by typical echo memory system limited by re-absorption \cite{Sangouard2007, Zhang2016}. While the required target parameters are not realistic for $\mathbb{E}=54\%$, 
we would like to address the question of how to achieve reasonably  high $\mathbb{E}$ with an optimization of  resonant thickness? As a reference a flat comb with ($S$, $\overline{\xi}$) = (50, 8) results in $\mathbb{E}=47.7\%$.

In order to match NFC and the spectrum of an  input Gaussian pulse, we  let $\xi_n$  proportional to a normalized Gaussian distribution
\begin{equation}
P\left( n, k \right) = \frac{ Exp  \left\lbrace      -\left[ \frac{1}{2} k\tau_p \left( n-\frac{M+1}{2}\right)  S \Gamma\right]^2        \right\rbrace  }{\sum\limits_{m=1}^{M} Exp  \left\lbrace      -\left[ \frac{1}{2} k\tau_p \left( m-\frac{M+1}{2}\right)  S \Gamma\right]^2        \right\rbrace    },
\end{equation}
where the total target number  $M\in$ odd.
%$m\in \left\lbrace -2, -1, 0, 1, 2 \right\rbrace  $ for correspondingly numbered targets in Fig.~\ref{fig1}(a).
%
This spectral shaping can be achieved by controlling either the isotope concentration or the target thickness during crystal growth.
The modified comb structure in Fig.~\ref{fig1}(b) is given  by multiplying the flat comb \cite{Zhang2018} and $P\left( n, k \right)$, where parameter $k$ is used to adjust the overall width of NFC, e.g., $k=0$ for a flat comb and $k=1$ for a NFC width equals to the bandwidth of the incident pulse.
The analytical solution of the output signal due to spectral shaping reads (see Supplementary Material for derivation)
\begin{widetext}
\begin{eqnarray}
\Omega_M\left( t, L_M\right) 
&=& \frac{\Omega_0\sqrt{\pi}\tau_p}{2T}\sum\limits_{l=-\infty}^{\infty}Exp\left[ 
-\left( \frac{l\pi\tau_p}{2T}\right)^2
+i l\pi\frac{t-\tau_i}{T}
-2i \xi\Gamma
\sum\limits_{n=1}^{M}\frac{P\left( n, k \right)}{-l\frac{\pi}{T}-\left( n-\frac{M+1}{2}\right) S\Gamma+i\frac{\Gamma}{2}}
\right], \label{eq11} 
\end{eqnarray}
\end{widetext}
where $\xi=M \overline{\xi}=\sum_{n=1}^M \xi_n$ is the total resonant thickness, and the introduced $T$ must be long enough to cover the complete output signal.

%%%%%%%%%%%%%%%%%%%%%%%%%%%%%%%%%%%%%%%%%%%%%%%%%%%%%%%%%%%%%%%%%%%%%%%%%
\begin{figure}[b]
\vspace{-0.4cm}
  \includegraphics[width=0.44\textwidth]{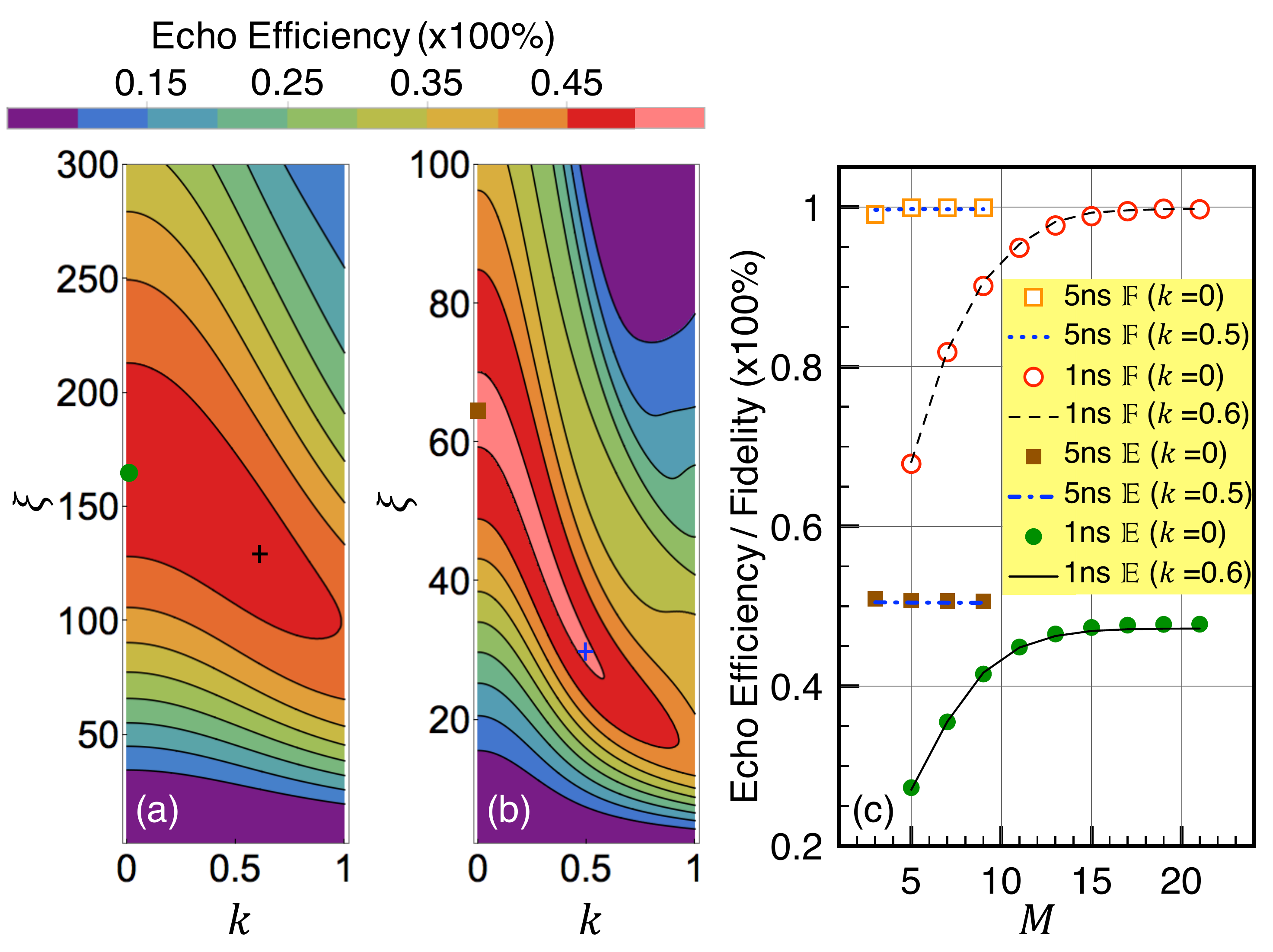}
  \caption{\label{fig2}
(Color online) ($k, \xi $)-dependent echo efficiency $\mathbb{E}$ based in analytical solution Eq.~(\ref{eq11}) for (a) ($\tau_p$, $S$, $M$) = (1ns, 50, 21) and (b) ($\tau_p$, $S$, $M$) = (5ns, 50, 9).
Green dot, black cross, brown-filled square and blue cross indicate the parameter sets 
($k, \xi, M $) = (0, 166, 21), 
(0.6, 129, 21), 
(0, 64, 9) and 
(0.5, 30, 9), 
respectively, used in (c) to show the reduction of total resonant thickness due to spectral shaping.
(c) $M$-dependant $\mathbb{E}$ and fidelity $\mathbb{F}$. 
For $\tau_p =5$ns, orange squares and brown-filled squares (blue-dotted line and blue-dashed-dotted line) respectively depict $\mathbb{F}$ and $\mathbb{E}$ for $k=0$ ($k=0.5$).
For $\tau_p =1$ns, red circles and green dots (black-dashed line and black-solid line) respectively illustrate $\mathbb{F}$ and $\mathbb{E}$ for $k=0$ ($k=0.6$).  
}
\end{figure}
%%%%%%%%%%%%%%%%%%%%%%%%%%%%%%%%%%%%%%%%%%%%%%%%%%%%%%%%%%%%%%%%%%%%%%%%%

The consistency between Eq.~(\ref{eq11}) and numerical solution has been carefully checked (see Supplementary Material).
Fig.~\ref{fig2} shows the ($k$, $\xi$)-dependent echo efficiency based in Eq.~(\ref{eq11}) and demonstrates the general effect of spectral shaping technique. An unperturbed single target  can  be utilized to control only a tiny Fourier fraction of a broadband input pulse whose bandwidth  is much larger than the natural linewidth of the absorber.
The use of very high $\xi$ may ease the broadband difficulty but will degrade $\mathbb{F}$ due to dispersion \cite{Liao2014b}. 
Spectral shaping technique in NFC will lower the needed $\xi$ and provides a solution to above problem.
In Fig.~\ref{fig2}(a) we demonstrate a broadband echo efficiency by using ($\tau_p$, $S$, $M$) = (1ns, 50, 21), i.e., an input bandwidth of 141$\Gamma$.
It plainly exhibits, on one hand, the $k$-dependent improvement of $\mathbb{E}$, e.g.,  $\mathbb{E}$ is prominently raised from 30\% to 44\% when $k$ = 1 is introduced for $\xi=80$.
On the other hand, the $k$-induced reduction of needed $\xi$ for achieving some $\mathbb{E}$  can be observed for all contours. 
Fig.~\ref{fig2}(b) shows another prominent example for ($\tau_p$, $S$, $M$) = (5ns, 50, 9), namely, a narrower input bandwidth of  28$\Gamma$. The required $\xi$ for $\mathbb{E}>50\%$ significantly decreases from 60 for $k=0$ to 26 for $k=0.5$.
One can see that $P\left( n, k>0 \right)$-modified NFC not only preserves the echo performance but also offers the advantage of the reduction of required $\xi$. Comparing shaped NFC with flat comb, maximum 22\% and 53\%  reduction of $\xi$ are observed  for $\tau_p=1$ns and $\tau_p=5$ns, respectively, whose parameters are indicated by green dot, black cross, brown-filled square and blue cross in Fig.~\ref{fig2}(a) and Fig.~\ref{fig2}(b).
The $P\left( n, k \right)$-modified NFC therefore optimizes the use of resonant thickness in contrast to a flat comb.
In view of Fig.~\ref{fig1}(b),  $\mathbb{E}$ of a narrow band input is mainly determined by the central resonant line and $\Delta = \pm S\Gamma$ side bands, because the spectral intensity of incident photons in this range is high. Including high order side bands mainly leads to even higher $\mathbb{F}$ because more frequency components are involved in the re-phasing process. One can therefore remove targets  contributing $\Delta > S \Gamma$ side bands with negligible change in $\mathbb{E}$. 
Along this line, in Fig.~\ref{fig2}(c) we discuss the reduction of target number $M$  via finding maximum $\mathbb{E}$ in the domain of  $0\leq \overline{\xi}\leq 15$ for each $M$, and calculate the corresponding fidelity $\mathbb{F}$ \cite{Chen2013, Liao2014b, Zhang2018}, where
\begin{equation}
\mathbb{F}=\frac{\vert \int_{t_1}^{t_2}\Omega_1^{\ast}\left( t - \tau_e, 0\right) \Omega_M\left( t, L\right)    dt \vert^2}{ \left[   \int_{-\infty}^{\infty} \vert \Omega_1\left( t , 0\right)  \vert^2   dt \right] \left[  \int_{t_1}^{t_2}\vert \Omega_M\left( t , L\right)  \vert^2   dt \right]}. \nonumber
\end{equation}
Brown-filled squares and green dots respectively depict $M$-dependent  maximum $\mathbb{E}$ for $\tau_p=5$ns and for $\tau_p=1$ns using flat comb ($k=0$). Orange squares and red circles are their fidelity.
The 5ns echo performance already reaches $\mathbb{E}=51\%$ and $\mathbb{F}=99\%$ as $M=3$.
However, more targets ($M>11$) are needed for $\tau_p=1$ns to achieve $\mathbb{E}>47\%$ and $\mathbb{F}>97\%$ due to its larger bandwidth than both $\Gamma$ and comb spacing. 
To manifest the advantage of spectral shaping technique, we show another set of  $\mathbb{E}$  and $\mathbb{F}$ for $\tau_p=1$ns  using $k=0.6$  with black-solid line and black-dashed line, respectively. 
For $\tau_p=5$ns, blue-dotted line and blue-dashed-dotted line respectively illustrate $\mathbb{F}$ and $\mathbb{E}$ using $k=0.5$.
It deserves to emphasize the versatile potential of spectral shaping technique with which one can essentially engineer $P\left( n, k \right)$ to cooperate with the input spectrum or to get desired output echo signals.

\begin{figure}[b]
\vspace{-0.4cm}
  \includegraphics[width=0.44\textwidth]{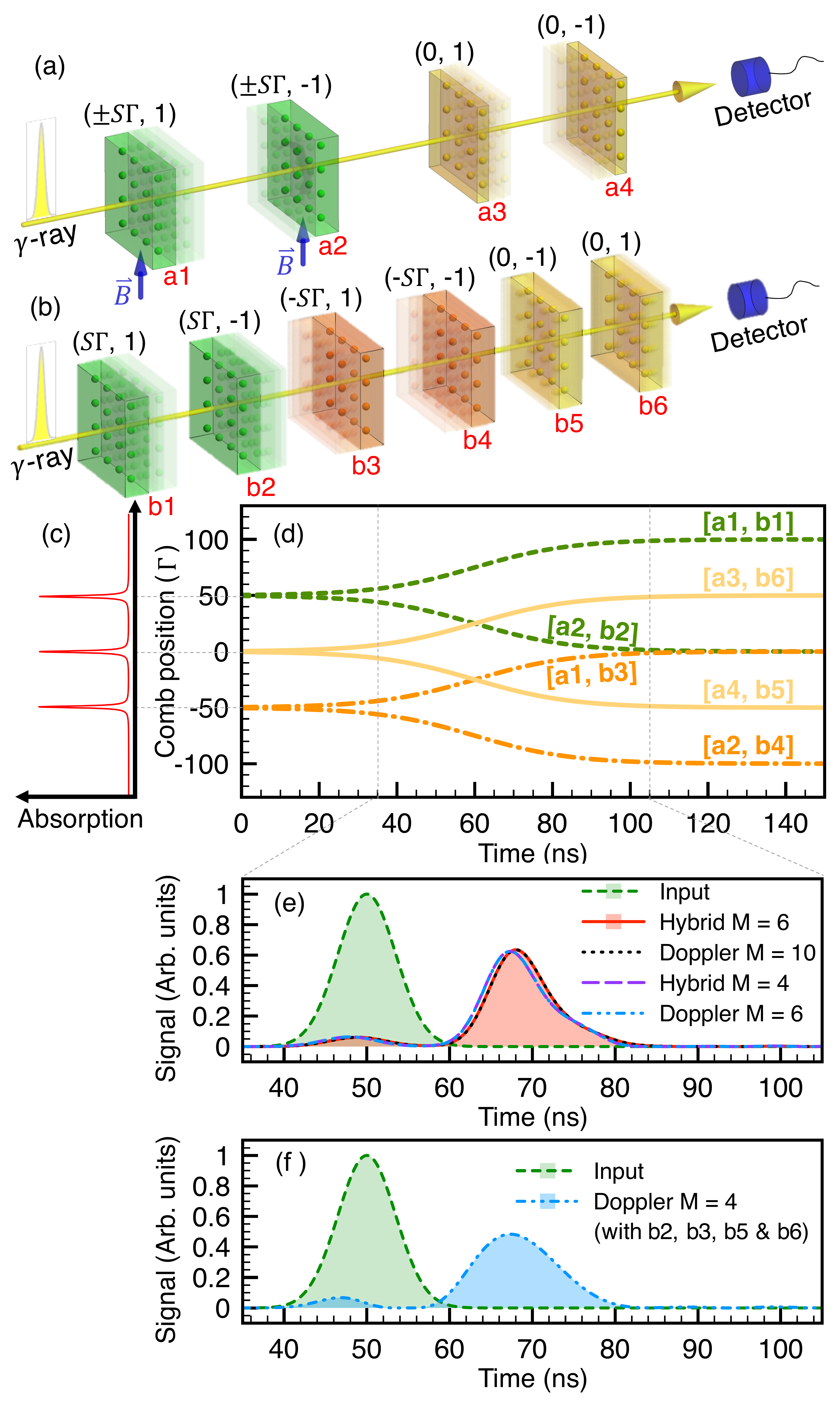}
  \caption{\label{fig3}
(Color online) 
%Improvement of echo efficiency by dynamical control over nuclear frequency comb. 
dynamical nuclear frequency comb system undergoes (a) hybrid shift (hyperfine splitting and Doppler shift) and (b) pure Doppler shift. Blue upward arrows illustrate applied magnetic fields $\vec{B}$, and coloured cuboids with afterimage demonstrate accelerated targets, which result in
(c) initial comb and
(d) dynamical splitting of nuclear frequency comb due to acceleration.
(e) dynamical-comb-improved echo efficiency of 67\% for an input 7ns $\gamma$-ray pulse (green-dashed-filled line).
Red-solid-filled, black-dotted, purple-long-dashed and blue-short-dashed-dotted-dotted line respectively depict targets with hybrid shift for $M=6$, pure Doppler shift for $M=10$, hybrid shift for $M=4$, and pure Doppler shift for $M=6$.
(f) echo efficiency of 66\% due to pure Doppler shift for $M=4$ with only targets \textbf{b2}, \textbf{b3}, \textbf{b5} and \textbf{b6}.    
Above each target, $\left( \pm\Delta_g \pm\Delta_e, \epsilon\right)$ is labelled in (a), and $\left( \Delta_D, \epsilon\right)$ in (b).
  }
\end{figure}

We now turn to address the issue of how to achieve $\mathbb{E} > 54\%$.
In a stationary comb system, $\mathbb{E}$ is mainly restricted by the re-absorption during phase revival and pulse propagation \cite{Sangouard2007, Zhang2018}. 
It is therefore critical to find a way to ease the re-absorption of emitted photons.
For this we invoke the dynamical splitting of NFC illustrated in Fig.~\ref{fig3}(a).
Initially, four targets are stationary.  
Magnetic field $\vec{B}$ (blue upward arrows) is applied to only targets \textbf{a1} and \textbf{a2} (green cuboids) whose nuclei experience hyperfine splitting $\Delta_g+\Delta_e =  S\Gamma$. Targets \textbf{a3} and \textbf{a4} (yellow cuboids) are not magnetized and contribute unperturbed resonance to NFC.
Fig.~\ref{fig3}(c) illustrates the initial absorption peaks at zero and $\pm S\Gamma$ induced by constant hyperfine splitting.
Subsequently, targets \textbf{a1} and \textbf{a3} (\textbf{a2} and \textbf{a4}) are backward (forward) accelerated and experience corresponding time-dependent Doppler shift $\Delta^a_D(t)$.
The total energy shift results from both mechanical acceleration and magnetic field, namely, hybrid shift, is described by
$\Delta_{31}=\Delta_g+\Delta_e+\Delta^a_D(t)$ and $\Delta_{42}=-\left( \Delta_g+\Delta_e\right) + \Delta^a_D(t)$, where
$
\Delta^a_D(t) = 0.5 \epsilon S\Gamma\left\lbrace  1+ \tanh\left[\left( t-\tau_d\right) /\left( 0.25 b_d\right)  \right] \right\rbrace.
$
Here $b_d$ describes the raising time of the target velocity  and $\tau_d$ is the middle time of the acceleration. 
$\epsilon =  -1$, $0$, or $1$ respectively depicts the forward acceleration (red Doppler shift), stationary or  backward acceleration (blue Doppler shift).
For a given comb spacing $S\Gamma$, the terminal velocity of an accelerated target $v= c S\Gamma/ \omega$ where $\omega$ is the unperturbed transition angular frequency, and during the period of $\left[\tau_d-b_d/2, \tau_d+b_d/2 \right]$ a NFC tooth at $m\Gamma$ is moving to $\left( m +\epsilon \right) \Gamma$. 
Fig.~\ref{fig3}(d) depicts the NFC evolution due to above-mentioned time sequence. The contributors to each line are labelled, e.g., the top green-dashed-bent-upward line is given by target \textbf{a1}  in Fig.~\ref{fig3}(a) and by target \textbf{b1}  in Fig.~\ref{fig3}(b). Each NFC line splitting originates from a corresponding pair of targets under opposite  acceleration.
%
%
%%%%%%%%%%%%%%%%%%%%%%%%%%%%%%%%%%%%%%%%%%%%%%%%%%%%%%%%%%%%%%%%%%%%%%%%%
%%%%%%%%%%%%%%%%%%%%%%%%%%%%%%%%%%%%%%%%%%%%%%%%%%%%%%%%%%%%%%%%%%%%%%%%%
Alternatively, as illustrated in Fig.~\ref{fig3}(b), one can invoke pure mechanical motion with
$\Delta_{31}=\Delta_{42}=\Delta_D+\Delta_D^a(t)$
to equivalently construct above dynamical splitting of NFC. 
$\Delta_D$ is given by the initial velocity of each target, and $\Delta_D^a(t)$ is also introduced by acceleration.
Green, orange and yellow cuboid  respectively demonstrate target with $\Delta_D = S\Gamma$, $-S\Gamma$ and zero. 

\begin{figure}[b]
\vspace{-0.4cm}
  \includegraphics[width=0.38\textwidth]{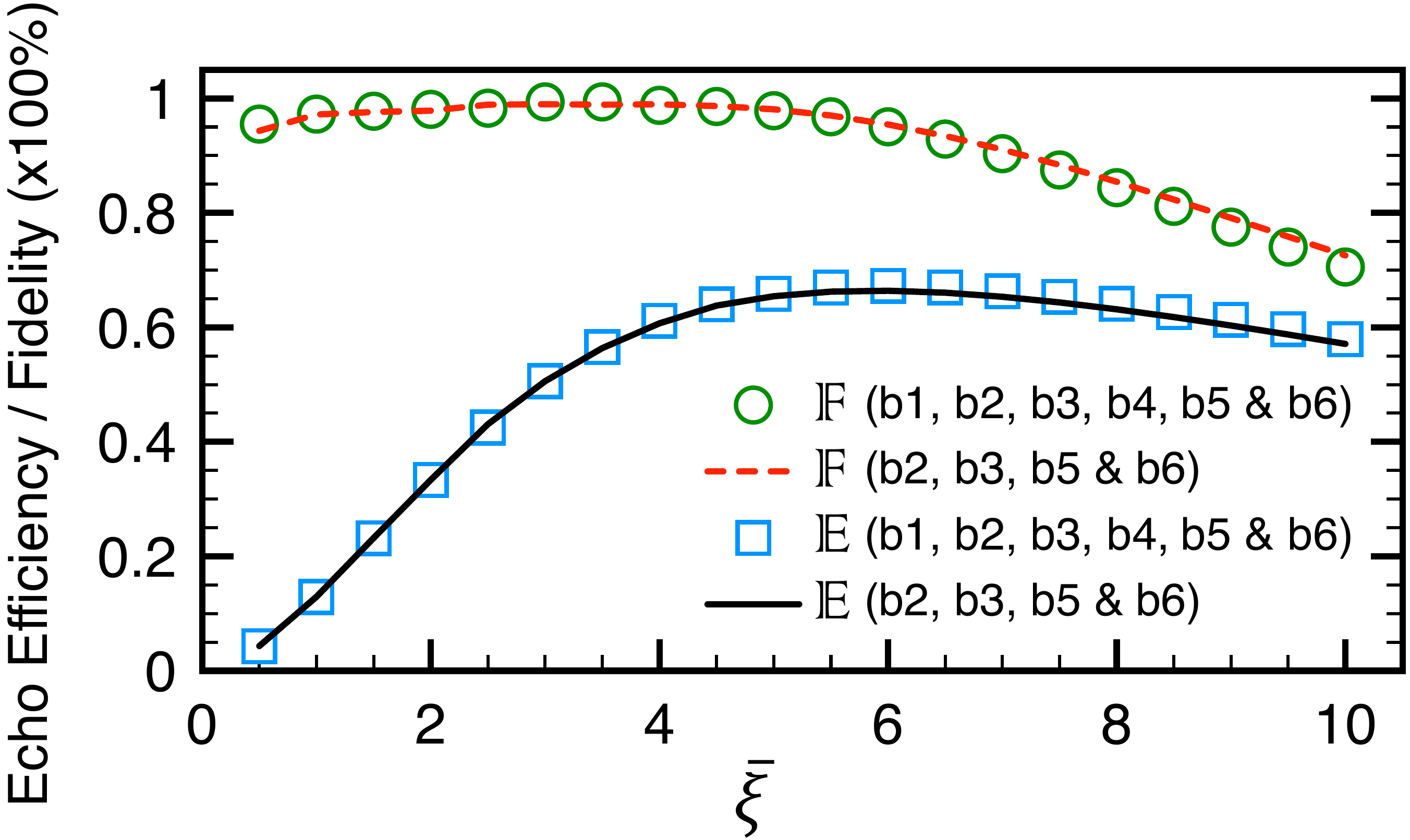}
  \caption{\label{fig4}
(Color online) comparison of $\gamma$-ray echo between  with and without target \textbf{b1} and \textbf{b4} in Fig.~\ref{fig3}(b). Blue squares and black solid line (green circles and red-dashed line) illustrate efficiency $\mathbb{E}$ (fidelity $\mathbb{F}$) with and without target \textbf{b1} and \textbf{b4}, respectively.
}
\end{figure}

Three purposes of utilizing $\Delta^a_D(t)$ are 
(i) the absorption of the incident  pulse can be enhanced by the part of comb teeth drifting inward the central resonant line; 
(ii) the comb is modulated so that the re-absorption condition is avoided during the whole emission process because emitters become off-resonant to re-absorbers due to time-dependent Doppler shift.
Also, the input photons are distributed in different target depending on their frequency.
The dynamical comb is therefore an  analogy to gradient echo memory (GEM) \cite{Hetet2008};
(iii) the frequency crossing at around $t=$ 60ns corresponds to the reversing of the spatial gradient and causes the echo emission \cite{Hetet2008}.
With the numerical solution of Eq.~(\ref{eq1}-\ref{eq3}) (see Supplementary Material for numerical method), Fig.~\ref{fig3}(e) demonstrates the echo signal due to above dynamical NFC system  illustrated  in Fig.~\ref{fig3}(d).
An optimization of $\Delta_D^a(t)$ shows that $\mathbb{E}$ is significantly raised to 67\% with $\mathbb{F}=96\%$ for an input 7ns  pulse (green-dashed-filled line) when using ($\tau_d$, $b_d$) = (60ns, 100ns). 
Four cases are presented 
by red-solid-filled line for hybrid shift and $M=6$ [six  targets with $\left( \Delta_g+\Delta_e, \epsilon\right) = \left( 2 S\Gamma, 1\right)$, $\left( 2 S\Gamma, -1\right)$, $\left( S\Gamma, 1\right)$, $\left( S\Gamma, -1\right)$, $\left(0, 1\right)$ and $\left(0, -1\right)$]; 
by black-dotted line for Doppler shift and $M=10$ [ten  targets with $\left( \Delta_D, \epsilon\right) = \left(2 S\Gamma, -1\right)$, $\left(2 S\Gamma, 1\right)$, $\left(-2 S\Gamma, -1\right)$, $\left(-2 S\Gamma, 1\right)$, $\left( S\Gamma, -1\right)$, $\left( S\Gamma, 1\right)$, $\left( -S\Gamma, -1\right)$, $\left( -S\Gamma, 1\right)$, $\left(0, -1\right)$ and $\left(0, 1\right)$];
by purple-long-dashed line for hybrid shift and $M=4$ illustrated in Fig.~\ref{fig3}(a); 
by blue-short-dashed-dotted-dotted  line for Doppler shift and $M=6$ depicted in Fig.~\ref{fig3}(b).
Other parameters are ($\tau_p$, $S$, $k$, $v$) = (7ns, 50, 0, 4.83mm/s), $\overline{\xi}=11.2$ for hybrid shift and $\overline{\xi}=5.6$ for pure Doppler shift.
One can perceive the equivalence of two systems by seeing the coincidence of both echo signals.
We also use a 5ns pulse as an input and obtain $\mathbb{E}=65\%$ and $\mathbb{F}=87\%$. 
In the light of above purpose (iii) and Fig.~\ref{fig3}(d),  target \textbf{b1} and \textbf{b4} should contribute very limited fraction to the echo. As depicted in Fig.~\ref{fig3}(f), we remove target \textbf{b1} and \textbf{b4} from the system of Fig.~\ref{fig3}(b)  and get an echo signal with $\mathbb{E}=66\%$ and $\mathbb{F}=95\%$ for a 7ns input pulse, which is very close to what we show in Fig.~\ref{fig3}(e). 
Fig.~\ref{fig4} shows the comparison of $\overline{\xi}$-dependent $\mathbb{E}$ and $\mathbb{F}$ between with and without target \textbf{b1} and \textbf{b4} in Fig.~\ref{fig3}(b). The consistency suggests the analogy between our dynamical comb and GEM. However, its inhomogenous energy broadening  is implemented by temporal variation rather than by only spatial gradient. This distinction allows dynamical splitting scheme using only few targets  to achieve $\mathbb{E}> 54\%$ comparing to our previous proposal of stepwise GEM \cite{Zhang2018}, and significantly reduces the complexity of implementation.
%
%%%%%%%%%%%%%%%%%%%%%%%%%%%%%%%%%%%%%%%%%%%%%%%%%%%%%%%%%%%%%%%%%%%%%%%%
%%%%%%%%%%%%%%%%%%%%%%%%%%%%%%%%%%%%%%%%%%%%%%%%%%%%%%%%%%%%%%%%%%%%%%%%%
The dynamical splitting method can be applied to other wavelength region and  provides a general way to achieve higher echo efficiency than the theoretical maximum \cite{Sangouard2007, Zhang2018}. As depicted in Fig.~\ref{fig3}(d), each target is accelerated to few mm/s within  100ns for the implementation of the present scheme. This is feasible \cite{Zhang2018} because reversing velocity in an even shorter  interval of about 10ns  had be demonstrated \cite{Helisto1991, Tittonen1993, Shakhmuratov2011, Shakhmuratov2013}. Moreover, controllable few ns $\gamma$-ray source generating incident pulse  had be demonstrated \cite{Vagizov2014} and proposed \cite{Wang2018}.

In conclusion  we have demonstrated two types of spectral control over $\gamma$-ray echo, 
which break the theoretical maximum echo efficiency of 54\% \cite{Sangouard2007, Zhang2018}. 
Our results lay the foundation of  developing  $\gamma$-ray photon memory in 10keV regime  \cite{Zhang2018}.
The present scheme also suggests a $\gamma$-ray echo device providing tunable time delay, which is typically difficult by using exquisite $\gamma$-ray optics.

C.-J. Y., P.-H. Lin and W.-T. L. are supported by the Ministry of Science and Technology, Taiwan (Grant No. MOST 107-2112-M-008-007-MY3 and Grant No. MOST 107-2745-M-007-001-). 
W.-T. L. is also supported by the National Center for Theoretical Sciences, Taiwan.

%%%%%%%%%%%%%%%%%%%%%%%%%%%%%%%%%%%%%%%%%%%%%%%%%%%%%%%%%%%%%%%%%%%%%%%%%

%%%%%%%%%%%%%%%%%%%%%%%%%%%%%%%%%%%%%%%%%%%%%%%%%%%%%%%%%%%%%%%%%%%%%%%%%
\bibliographystyle{apsrev}
\bibliography{NFSBS2A}
%%%%%%%%%%%%%%%%%%%%%%%%%%%%%%%%%%%%%%%%%%%%%%%%%%%%%%%%%%%%%%%%%%%%%%%%%%%%%%%%
%%%%%%%%%%%%%%%%%%%%%%%%%%%%%%%%%%%%%%%%%%%%%%%%%%%%%%%%%%%%%%%%%%%%%%%%%%%%%%%%

\end{document}